\documentclass{./styles/svproc}
\usepackage{url}
\usepackage{cite}
\usepackage{amsmath,amssymb,amsfonts}
\usepackage{algorithmic}
\usepackage{graphicx}
\usepackage{textcomp}
\usepackage{xcolor}
\usepackage{subcaption}

\begin{document}
\mainmatter              
\title{Robust and Efficient AI-Based Attack Recovery in Autonomous Drones\\
}
%
%
\author{Ivar Ekeland\inst{1} \and Roger Temam\inst{2}
Jeffrey Dean \and David Grove \and Craig Chambers \and Kim~B.~Bruce \and
Elsa Bertino}
\author{Diego Ortiz Barbosa\inst{1} \and Luis Burbano\inst{1} \and Siwei Yang\inst{1} \and Zijun Wang\inst{1} \and \\Alvaro A. Cardenas\inst{1} \and Cihang Xie\inst{1}
\and Yinzhi Cao \inst{2}
}
%
%
\institute{University of California Santa Cruz, Santa Cruz CA 95064, USA\\
\and
Johns Hopkins University, Baltimore MD 21218, USA}

\maketitle              

\begin{abstract}
We introduce an autonomous attack recovery architecture to add common sense reasoning to plan a recovery action after an attack is detected. We outline use-cases of our architecture using drones, and then discuss how to implement this architecture efficiently and securely in edge devices. 
\end{abstract}
\section{Introduction}

Autonomous drones or self-driving vehicles are vulnerable to various attacks, such as physical interference affecting sensor readings~\cite{yan2020sok}, actuation signals~\cite{dayanikli2022physical}, GPS spoofing~\cite{sathaye2022experimental}, etc. Such security lapses can cause dangerous consequences in the physical world, such as vehicle crashes~\cite{car} or navigation errors~\cite{rutkin2013spoofers} that may steer our autonomous vehicle into enemy territory or away from its mission.

To protect these systems, researchers have developed several tools for preventing, detecting, and recovering from attacks. Automatic recovery, the last of these steps, plays a significant role for drones and other autonomous vehicles because if they are attacked, they need to recover quickly to prevent accidents such as crashing or harming humans. 

Real-time attack recovery solutions are mainly based on the \textbf{simplex architecture}, which consists of two \emph{different} controllers~\cite{dash2021pid, zhang2020real,garg2022control}: One is a nominal controller optimized for performance but without safety guarantees. If an attack is detected, we switch from the nominal controller to the \textit{recovery controller}, a controller that changes the objective of the mission to perform a safety maneuver. These recovery controllers can try to steer the drone to a safe area, even when signals are partially compromised.

\begin{figure}[htb]
    \centering
    \begin{subfigure}[b]{.32\linewidth}
    \includegraphics[width=\linewidth]{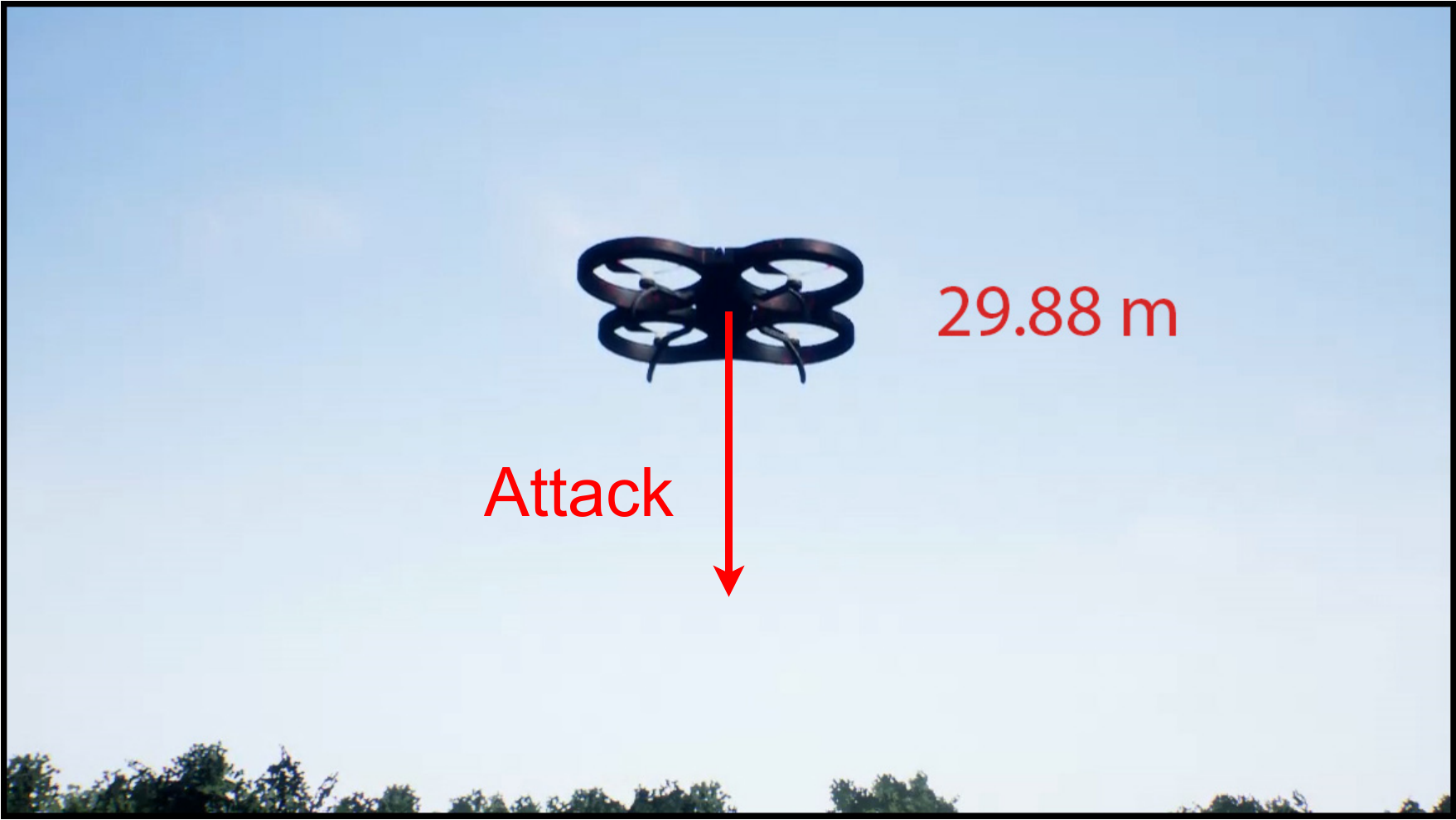}
    \vspace*{-15pt}
    \caption{Hovering.}\label{fig:drone:normal}
    \vspace*{5pt}
    \end{subfigure}
    \begin{subfigure}[b]{.32\linewidth}
    \includegraphics[width=\linewidth]{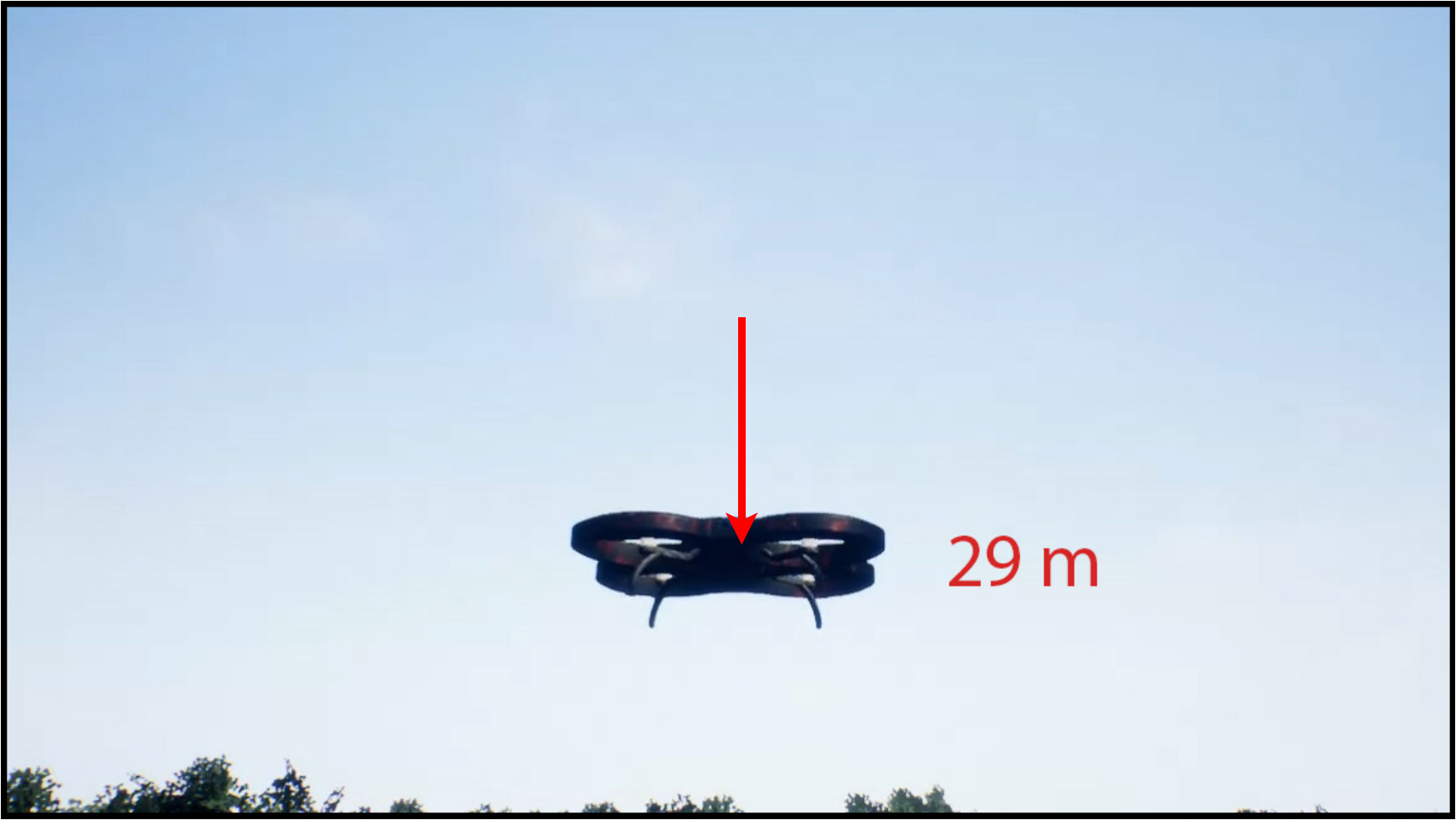}
    \vspace*{-15pt}
    \caption{Attacked.}\label{fig:drone:attacked}
    \vspace*{5pt}
    \end{subfigure}
    \begin{subfigure}[b]{.32\linewidth}
    \includegraphics[width=\linewidth]{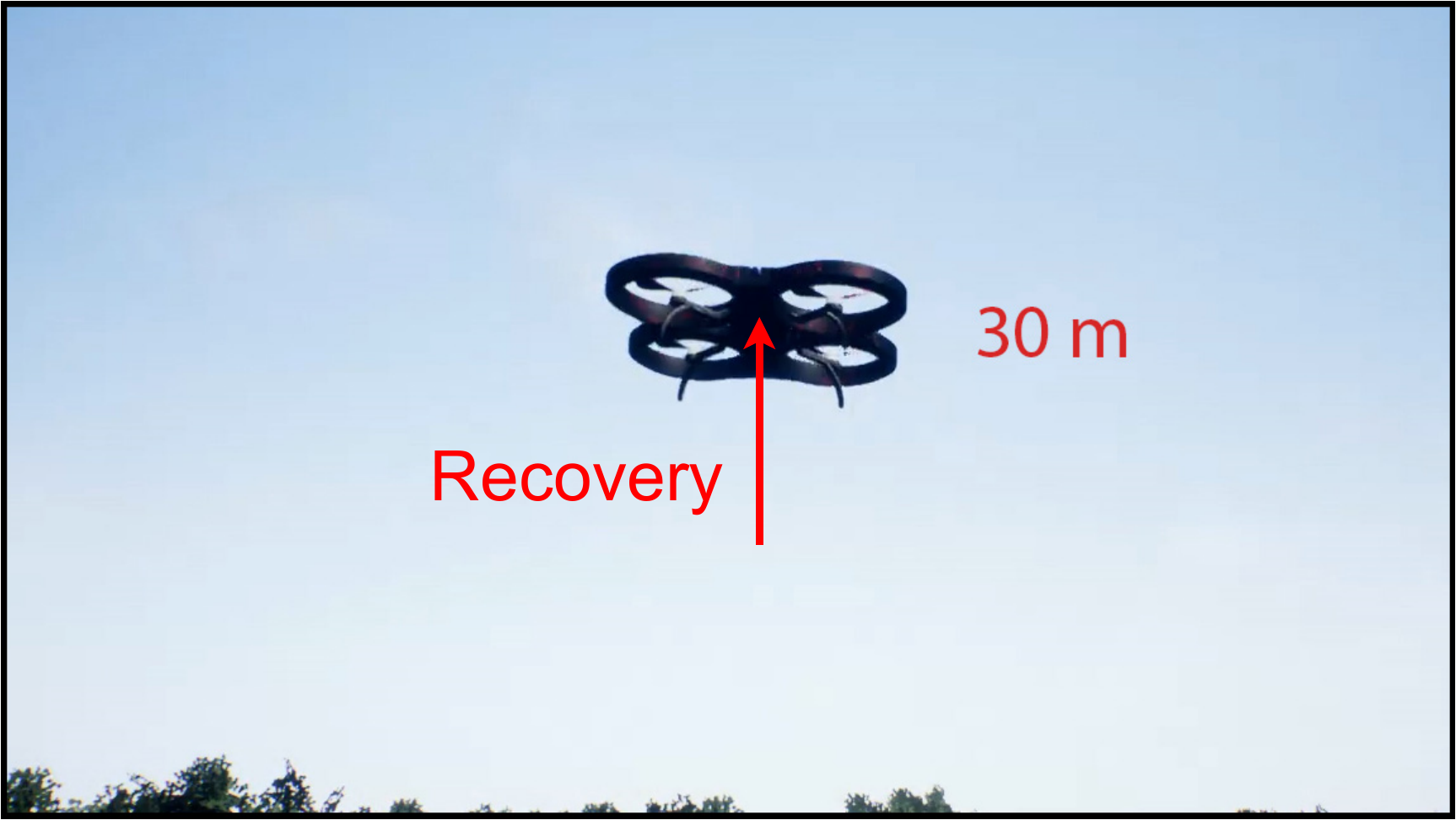}
    \vspace*{-15pt}
    \caption{Recovered.}\label{fig:drone:recovered}
    \vspace*{5pt}
    \end{subfigure} 
        \caption{A drone receives false GNSS information, forcing it to lower it's altitude. OPR detects this attacks and returns the drone to a safe altitude. } 
    \end{figure}

\subsection{Example}
We now illustrate how our work leverages this recovery controller to keep drones safe. Drones must perform different tasks, such as surveillance in adversarial environments, where attackers might want to land the drone without authorization or produce a crash. 

This example is motivated by the RQ-170 UAV incident. In particular, the government of Iran claims they used a cyber-attack to force a U.S. surveillance drone to land in Iranian space~\cite{jaffe2011iran,shane2011drone}. In this use case, an attack spoofs GPS signals to make the drone believe it is at a higher altitude than it really is (Figure~\ref{fig:drone:normal}). Without any defense, the drone will start descending and eventually land (Figure~\ref{fig:drone:attacked}). Our attack-recovery mechanism detects the attack (by looking at the inconsistency between control actions and GPS values) and then recovers its original (safe) position by creating virtual sensors: altitude predictions based on physical models (Figure~\ref{fig:drone:recovered}).

\begin{figure}[htb]
    \centering 
    \hspace{1em}\begin{subfigure}[b]{0.4\textwidth}
         \includegraphics[width=\textwidth]{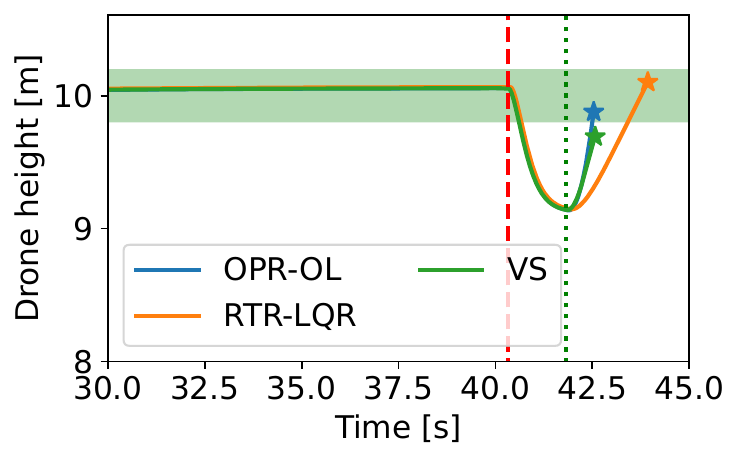}
         \caption{OPR-OL vs. baselines.}
         \label{fig:drone_timeseries}
    \end{subfigure}  
    \hfill
    \begin{subfigure}[b]{0.4\textwidth}
         \includegraphics[width=\textwidth]{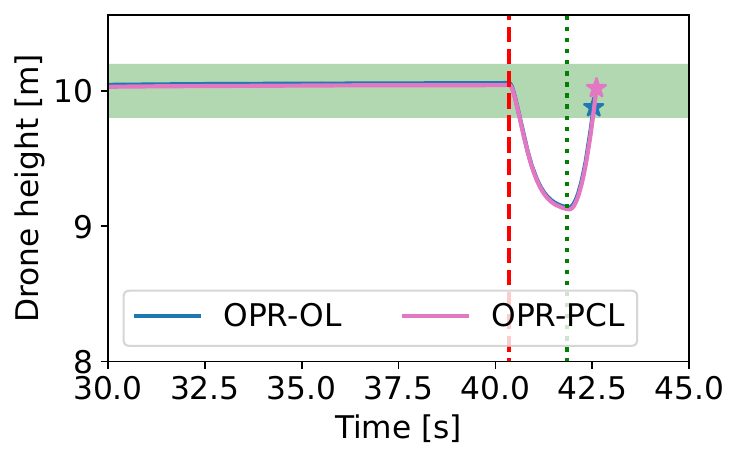}
         \caption{OPR-OL vs. OPR-PCL.}
         \label{fig:traxxas_opr_timeseries}
    \end{subfigure}    
             \caption{Our algorithm (OPR-OL) returns a drone to a safe height (green area) faster and more accurately than previous work. In addition, if we can filter out the malicious sensor and take the input from the remaining sensors, we obtain a Partially Closed Loop (OPR-PCL) algorithm that outperforms slightly our open loop model.} 
\end{figure}

We call our algorithm Optimal Probabilistic Recovery (OPR)~\cite{zhang2024fast} and we consider it as Open Loop (OL) if we assume that all sensors are compromised, or Partially Closed Loop (PCL) if we can detect the only signal attacked, and then consider the other sensors as trustworthy. Figure~\ref{fig:drone_timeseries} shows that OPR-OL recovers the drone faster than other baselines (Real-Time Recovery with the Linear Quadratic Regulator--RTR-LQR~\cite{zhang2020real} and Virtual Sensors--VS\cite{cardenas2011attacks}); and Figure~\ref{fig:traxxas_opr_timeseries} shows how information from non-compromised sensors (OPR-PCL) can improve recovery by landing in the middle of the target set. 

\begin{figure}[htb]
    \centering  
    \hfill\begin{subfigure}[b]{0.4\textwidth}
         \includegraphics[width=\textwidth]{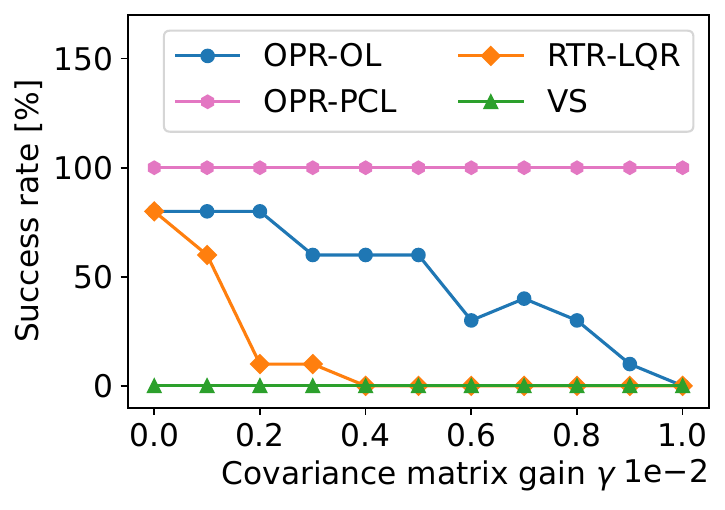}
         \vspace*{-15pt}
         \caption{Success rate}
         \label{fig:drone:success_noise}
    \end{subfigure}
    \hfill
    \begin{subfigure}[b]{0.4\textwidth}
         \includegraphics[width=\textwidth]{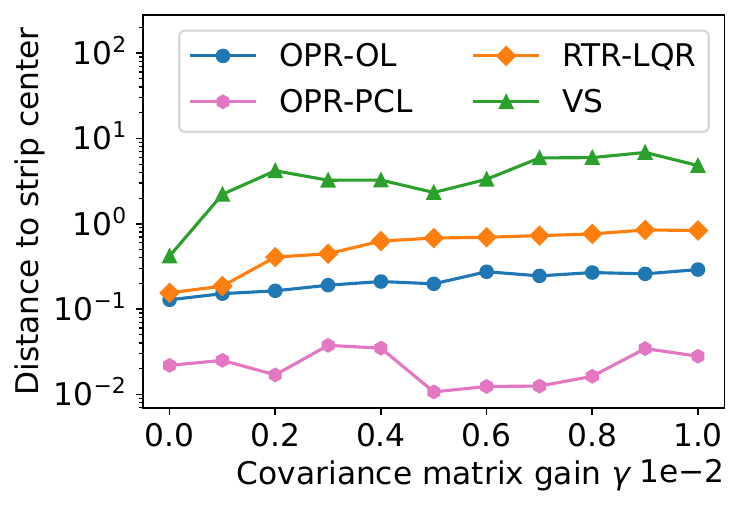}
         \vspace*{-15pt}
         \caption{Distance to strip center}
         \label{fig:drone:distance_noise2}
    \end{subfigure}    
    \caption{Success rate and average distance to the target set center with increasing noise for the drone.} 
    \label{fig:noise}
\end{figure}

OPR-OL and OPR-PCL also outperform the success rates of the baselines (how many attacks are recovered to the target--green--set, in Figure~\ref{fig:drone:success_noise}) and by the distance to the center of the desired target (Figure~\ref{fig:drone:distance_noise2}).

While these previous efforts can help prevent immediate safety risks, they still require mission planners to identify several parameters before a mission, such as safe destinations to go to (targets) after an attack is detected; and thus they are not adaptable to uncertain conditions and new attacks. In our ongoing work, we plan to address these limitations by leveraging advances in AI.

To make our AI-based attack recovery strategy useful and practical,  we argue that we need to solve the following research challenges:
\begin{itemize}
    \item Design of an AI recovery algorithm.
    \item Design of efficient and practical algorithms that can run on edge devices or on embedded systems by orchestration with an AI agent in the cloud. 
    \item Design attack-resilient AI agents that are not vulnerable to test-time adversarial example attacks. 
\end{itemize}

\section{Challenge 1: Autonomous Recovery}

The state-of-the-art automatic attack-recovery mechanisms described in the previous section do not work with dynamic and uncertain environments. For example, these previous methods need precomputed target safe areas where the recovery controller can take the system; however, if these sets are not preloaded in advance, or if the safe zones are not safe during sporadic periods of time, the automatic recovery mechanism will fail. 

\begin{figure}[ht]
    \centering  
         \includegraphics[width=.95\columnwidth]{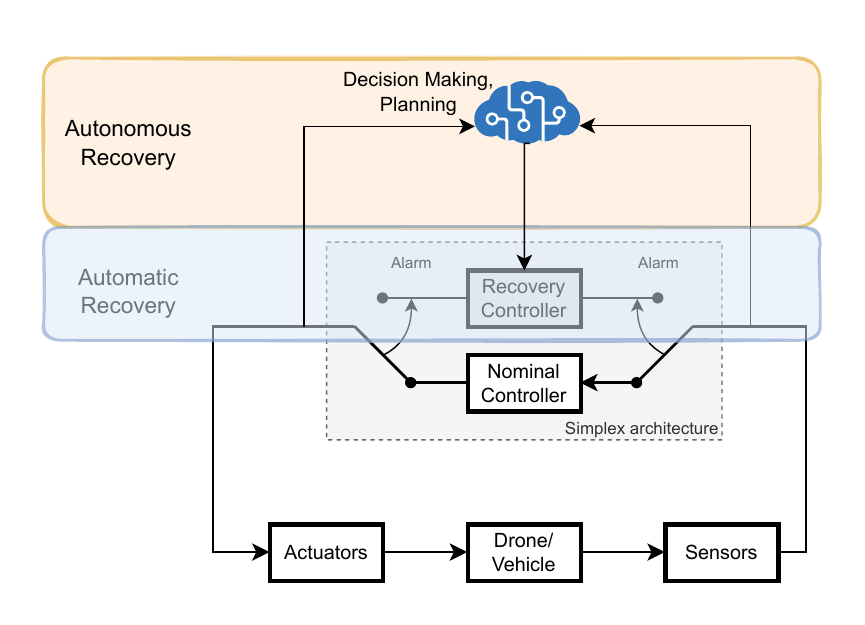}
         \caption{AI-Based Recovery.}
         \label{fig:recovery}
\end{figure}

As the cornerstone of a new era in AI, generative AI (GenAI) models such as Falcon2~\cite{malartic2024falcon2} and GPT-4~\cite{achiam2023gpt} promise to catalyze a profound transformation across numerous sectors of society, providing common sense reasoning in real time to adapt to uncertain and dynamic scenarios.
To address the limitations of previous attack recovery systems, we propose a GenAI-Based attack recovery mechanism. Our main insight is to have a hierarchical recovery strategy --- At the lower level we will use mathematical control-theory models based on the simplex architecture (as described in the previous section); At a higher level, we will design a generative AI recovery algorithm to provide a common-sense and adaptive recovery plan. Our concept is illustrated in Figure~\ref{fig:recovery}. 


To design an AI-based attack recovery, we need to solve several problems: (1) AI agents need to understand the state of the drone (or vehicle), identify risks, and create action plans. This requires encoding of the state of the physical world into a format that can be understood by the GenAI agent. (2) Identify safety zones dynamically as the mission progresses to give to the lower-level automatic controller,  (3) Have a long-term plan for recovering after reaching the target set (e.g., identify if the attack has stopped, when can we engage the nominal controller again, and when do we ask for help from a human operator or other agents).


In particular, we plan to extend our recent work \cite{zhang2024fast} with common sense reasoning to find safe target sets and maneuver toward them after we detect an attack. We define the target sets with two elements: 1) the closed form $T\in \mathcal{T}$, with $\mathcal{T}$ the set of possible forms, and 2) the parameters $\theta\in \Theta$, where $\Theta$ the set of all possible parameters. Note that $\theta$ depends on the form of the target set $T$. Then, we denote the set of valid parameters $\Theta$ for a target set form $T$ as $\Theta(T)$. For instance, for a drone with $n$ states, the target set can be a strip \cite{zhang2024fast}, where we can define that the drone state $x\in \mathbb{R}^n$ is between a range at the end of the recovery.
A strip is the intersection between two hyperplanes $T(\theta)=\{ x\in \mathbb R^n \,|\,\theta_1^Tx\geq \theta_2 \, \wedge\, \theta_1^Tx\leq \theta_3 \}$, where $\theta_1\in \mathbb R^n$, $\theta_2\in \mathbb R$ and $\theta_3\in \mathbb R$ are the target set parameters. Therefore, we can select $\theta_1$ to define the flying height of the target drone between $\theta_2$ and $\theta_3$.

 LLMs can produce the parameters $\theta$. For this, the LLM takes sensor information from the observation set $o \in \mathcal O$, the form of the set $T\in \mathcal{T}$, and contextual information such as environmental conditions $c \in \mathcal C$ to produce the target set parameters $\theta \in \Theta$. That is, the LLM becomes a function $F:\mathcal O\times \mathcal{T} \times \mathcal{C} \to \Theta$.

Using the LLM to define the target set parameters comes with several challenges. First, the LLM may output a target set that is not actually safe. Similarly, the target set may be infeasible; arriving at the target set the LLM generates may be impossible.  Therefore, we will work on a verifier mechanism that certifies the safety and feasibility of the target set. 

\begin{figure*}[ht]
    \centering  \vspace{-1em}
        \includegraphics[width=1\columnwidth]{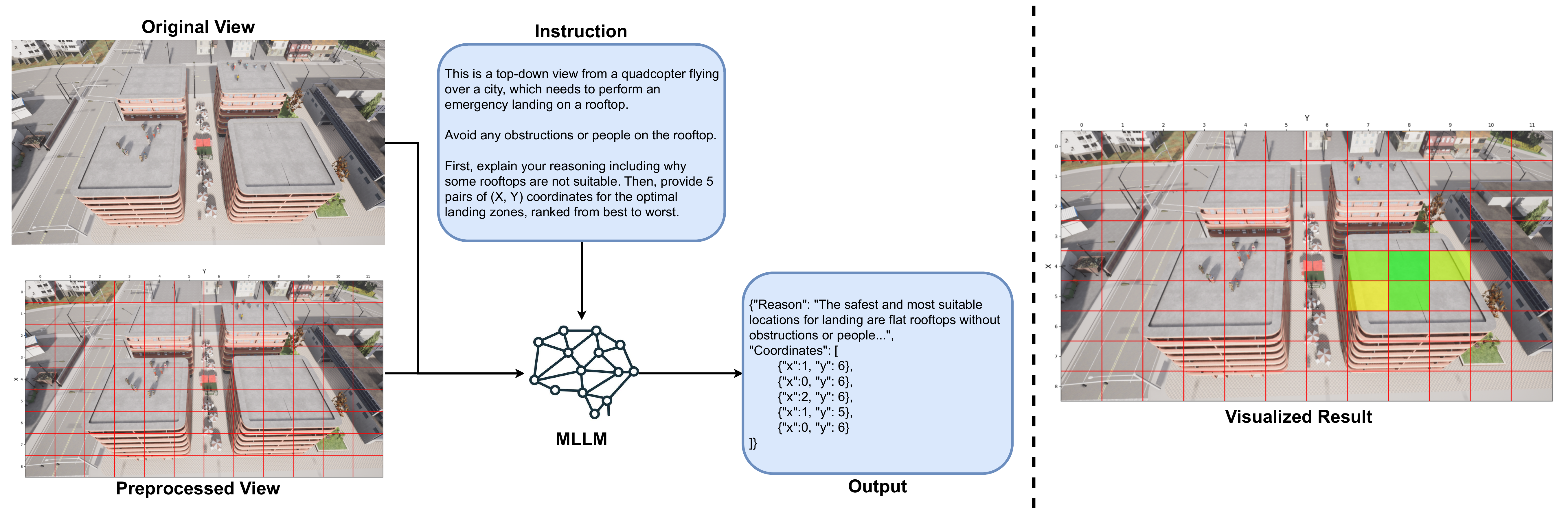}
         \caption{Multi-modal LLM (MLLM) evaluates the risks and ranks possible safe landing locations.}
         \label{fig:LLMAssistant}
\end{figure*}

Figure~\ref{fig:LLMAssistant} illustrates a use case of this methodology. After detecting an attack, we ask the LLM to identify a safe area where the drone can land (given the camera feed of the drone). The LLM must decide which of the four buildings the drone should land in. Two of those buildings are crowded with people, while the other two are empty. The LLM needs to identify that empty buildings are safer to land in than crowded ones. The drone's camera feed is first preprocessed with a Cartesian coordinate system added to make it easier to interpret the multi-modal LLM's output. The LLM then predicts several candidate safe landing zones based on both the original and preprocessed views. Grounded decoding is applied in the final stage of the LLM to ensure the output strictly follows the required format. Each predicted landing zone includes coordinates and a ``Reason'' section to improve prediction accuracy and interoperability.

To improve the reasoning process (and improve the prediction accuracy) of LLMs, we plan to test prompting techniques such as Chain-of-Thought, Self-Consistency, and Self-Reflection. Also, as LLMs may sometimes fail to recognize objects such as buildings and people in the images, a dedicated object detection/semantic segmentation model will be used to recognize objects and then color-code the objects in images as part of a preprocessing process, so that these objects can be easier for LLMs to recognize.

\section{Challenge 2: Efficient Edge and Device AI}

A critical challenge we face is the reduction of operational latency in GenAI applications. The success of drones in critical missions, such as immediate disaster response or high-speed surveillance operations, is highly dependent on their ability to process and respond to incoming information with minimal delay.

In responding to latency concerns, our aim is to tackle them with recent algorithmic efficiency proposals.

\begin{itemize}
     \item \textbf{Model Distillation:} This technique involves distilling a large language model into a more compact version while retaining the essential features necessary for robust performance. Following recent work \cite{b1}, our aim is to control the size of multimodal LLM under 0.2 billion parameters, ensuring rapidness without substantial loss in effectiveness.

     \item \textbf{Efficient Mobile Model Design:} Given that traditional transformer architectures exhibit quadratic computational complexity with respect to token length, exploring alternatives such as the Mamba / RWKV model \cite{b4,b5}, which offers linear complexity, is considered advantageous. This modification could significantly reduce computational demand, enabling quicker data processing \cite{b2,b3}.

    \item \textbf{Post-Training Quantization:} Transitioning from floating point precision (fp32 or fp16) to a highly quantization format such as int8 or even a binary version can substantially accelerate model operation \cite{b6}.
\end{itemize}

These three strategies can also be used together to further reduce model latency on edge devices, equipping drones with the capability to respond in real-time to diverse and dynamic environmental stimuli. Moreover, to build more capable multimodal LLMs, which requires navigating complex and varied real-world scenarios, we are exploring the following innovative approaches:

\begin{itemize} \item \textbf{Learning Every Signal:} To maximize the capabilities of multimodal LLMs, we plan to pioneer diverse tokenization methods aimed at integrating and processing a variety of signals. This strategic development is designed to build a coherent and multifaceted input landscape, encompassing different data types such as images, videos, textual and voice inputs from users, and radar signals. Our objective is to cultivate a robust input framework that significantly boosts the model's capacity to learn and adapt across the spectrum of data encountered in UAV operations.

\item \textbf{Reinforcement Learning with Human Feedback (RLHF):} We plan to incorporate human feedback into the training loop of our models. This can be achieved by engaging a human copilot who monitors and, if necessary, corrects the UAV's actions during operation. The corrective inputs provided by the human operator are used to reinforce and refine the model’s understanding and responses to real-world scenarios. By continuously evaluating and adjusting AI decisions with insights from experienced human experts, our goal is to significantly improve the decision-making capabilities of our systems, especially in complex environments where nuanced judgment and situational awareness are crucial. 
\end{itemize}

\section{Challenge 3: Robust GenAI-based Attack Recovery}

We also need strategies to enhance the robustness of GenAI systems to ensure that our recovery system is not abused by attackers. 

The high-level idea is that we apply randomized smoothing upon the inputs to a large language model and smooth its output, e.g., the decision on Drone's turning angles or flying directions.  Specifically, our method divides a given input prompt into several masked prompts with disjoint subsets of tokens.  Then, our method maps each token to an integer that indicates the index of the masked prompt. 
%
Then, our method assigns a token of the input prompt to the masked prompt. 
 Then, our method predicts an output for each masked prompt, takes a majority vote based on an epsilon ball of each output, and then takes the averaged output as the final result.  Since the method follows randomized smoothing, it will ensure that the output will not change much given an adversarial input. 

In the past, our previous work has studied different attacks against LLMs. We will use our attacks to evaluate the robustness of the proposed GenAI system.

\begin{itemize}

\item Jailbreaking Attack. Our jailbreaking attack searches for alternative tokens in replacing the filtered tokens in a given prompt while still preserving the prompt's semantics and the follow-up generated images.
 Our high-level idea relies on Reinforcement Learning (RL), which adopts agents to interact with text-to-image models' outputs and change the next action based on rewards related to two conditions: (i) semantic similarity, and (ii) success in bypassing safety filters.  Such RL agents not only solve the challenge of closed-box access to the text-to-image model but also minimize the number of queries as the reward function will guide the attack to find our adversarial prompts. 

\item Prompt Leaking Attack.   
 Our novel, closed-box prompt leaking attack is inspired by existing jailbreaking attacks~\cite{zou2023universal,wallace2019universal}. 
 It optimizes a query, which we call adversarial query, such that a target LLM application is more likely to reveal its system prompt when taking the query as input.  
Specifically, we formulate finding such an adversarial query as an optimization problem, which involves a dataset of {shadow system prompts} and a {shadow LLM}. For each shadow system prompt, we simulate a {shadow LLM application} that uses the shadow system prompt and the shadow LLM. 
Roughly speaking, the objective of our optimization problem is to find an adversarial query, such that the shadow LLM applications output their shadow system prompts as the responses for the adversarial query.

\end{itemize}
\section{Conclusions}
Future autonomous systems need to have fail-safe conditions that are adaptive to dynamical and unpredicted conditions. We propose an architecture for autonomous attack recovery and outline how to make it more efficient and secure. Our future work will evaluate this architecture methodologically and in realistic conditions.

\section*{Acknowledgments}
This material is based upon work supported in part by the Air Force Office of Scientific Research under award number FA9550-24-1-0015, and by the National Center for Transportation Cybersecurity and Resiliency (TraCR) (a U.S. Department of Transportation National University Transportation Center).
%
%
\bibliographystyle{splncs04}
\bibliography{references}
\end{document}